\documentclass[11pt,twoside]{article}
\usepackage{asp2010}

\resetcounters

\begin{document}

\title{Target for LOFAR Long Term Archive: Architecture and Implementation}
\author{Andrey Belikov,$^1$ Danny Boxhoorn,$^1$ Fokke Dijkstra,$^2$ Hanno A. Holties,$^3$ and Willem-Jan Vriend$^1$
\affil{$^1$ Kapteyn Astronomical Institute, University of Groningen, Landleven 12, 9747AB, Groningen, The Netherlands}
\affil{$^2$ Donald Smits Center for Information Technology, University of  Groningen, Nettelbosje 1, 9747AJ, Groningen, The Netherlands}
\affil{$^3$ ASTRON, Oude Hoogeveensedijk 4, 7991 PD, Dwingeloo, The Netherlands}}

\begin{abstract}
The LOFAR Long-Term Archive (LTA) is a multi-Petabyte scale data storage for the processed data of LOFAR telescope. We describe the adaptation of 
the WISE concept implemented by Target consortium for the LOFAR LTA and changes we introduced to it to accommodate LOFAR data. 
This paper describes an example of a new information system created on the basis of Astro-WISE for a wider range and scale of data. 
\end{abstract}

\section{Introduction}
\label{Introduction}

Astro-WISE~{\footnote{http://www.astro-wise.org}}~\citep{Astro-WISE} is an astronomical  scientific information system which was developed to host 
the data for the KIlo Degree Survey (KIDS\footnote{http://www.strw.leidenuniv.nl/$\sim$kuijken/KIDS/}). Astro-WISE 
combines data storage and data processing allowing the user to carry out intensive and in-depth data mining along with partial or 
full data reprocessing on the survey data.   

The development of the LOFAR Long Term Archive (LTA) put a completely new set of challenges for Astro-WISE which resulted in the creation of a 
new information system that will host the LTA. At the same time the requirements from LOFAR helped to improve Astro-WISE and make it more general adding 
new features to the system.  

The design of the LOFAR LTA was described in previous papers~\cite{LoWISE} and~\cite{LTA}. We describe the current state of LOFAR LTA, components, data storage and processing 
capacities and adaptation made to Astro-WISE architecture to host LTA data.

\section{LOFAR} 
\label{LOFAR} 

LOFAR\footnote{http://www.lofar.org}, the Low Frequency Array, is a key project in radioastronomy and a pathfinder instrument for the Square 
Kilometre Array~\footnote{http://www.skatelescope.org}. 
LOFAR was originally designed and constructed by ASTRON as an array in the Netherlands with baselines up to 100 km. Its capabilities are geared to key science projects on Deep Extragalactic Surveys, Transient Radio Phenomena and Pulsars, the Epoch of Reionoization, High Energy Cosmic Rays, Cosmic Magnetism, and Solar Physics and Space Weather. Now, the International LOFAR Telescope, mostly constructed by the end of 2010, and to be completed in 2011, has 40 stations in the Netherlands, plus at least 5 stations funded in Germany, and 1 each in France, Sweden, and the United Kingdom, that provide additional sensitivity and angular resolution on baselines extending to more than 1000 km. Future additions are envisaged.

The LOFAR system architecture is described in~\cite{ICSPC} and~\cite{IEEE2009}.

\section{LOFAR LTA Requirements}

The newly created LOFAR Long Term Archive is an information system which must fulfill the requirements of its users. These requirements include an implementation of a new 
data model, new data storage and adaptation to existing LOFAR data storage and data processing.

In 2011 LTA will store 1.5 PB of data and the capacity is expected to increase to 25 PB by the end of 2014. Because 
the research groups participating in LOFAR and the available storage and computing resources are distributed, the data 
must be stored in different geographical locations, while being accessible for retrieval and reprocessing by all LOFAR users. 
An important requirement is an open nature of the archive and data processing system. In practice this means that new 
groups of users (new projects) can be added to the LOFAR LTA with associated storage space and processing resources. 
They should be able to integrate their resources into the system with minimal efforts and resources spent on adaptation of the LOFAR LTA design.

\section{Adaptation of Astro-WISE}

Astro-WISE is an astronomical information system which was originally developed for the processing of optical wide-field images. Astro-WISE 
allows to process the data from the raw images up to catalogs keeping all dependencies and processing parameters in the system. The 
detailed description of Astro-WISE can be found in~\cite{JOGC}.

Astro-WISE is a distributed system with an infrastructure which is based on the principle of deploying independent nodes. Each node 
can contain a data storage (dataserver), metadata storage (relational DBMS), processing element (usually an HPC cluster) and numerous 
user services. All these components are optional for each site, as long as at least one of them is available elsewhere. 

The main challenge for the construction of a new information system for LOFAR on the basis of Astro-WISE is a conceptual difference between 
data storage and data processing for Astro-WISE and LOFAR. In the case of Astro-WISE all data processing from the raw images is done 
within the system, the raw data are coming in Astro-WISE with a complete data provenance. In the case of LOFAR the initial data processing 
is done in CEP and only processed data are archived in the Long-Term Archive. 

Astro-WISE was developed, tested and used for hundreds of Terabytes volume of data meanwhile LOFAR Long-Term Archive 
should be scaled up to tens of Petabytes.
The scale of the data storage and computing for LOFAR reaches the level of the Large Hadron Collider experiment (LHC).

After reviewing requirements described above, the architecture of the archive was changed from the federated architecture with independent nodes 
to a tier architecture similar to the architecture of LHC/EGEE~\citep{EGEEData}. 
Three tiers were introduced: 
\begin{itemize}
\item[-] Tier 0 is the Central Processing System (CEP) with a temporal off-line data storage which collects data from antennas, performs processing, and delivers the 
data to the Long Term Archive; 
\item[-] Tier 1 consists of a number of data storage nodes and computing nodes deployed in Groningen (Donald Smits Computing Center, CIT) or at any location 
where the research group participating in the project can provide a facility for data storage. A node of Tier 1 can be an Astro-WISE 
dataserver, a distributed filesystem node (GPFS) or a grid based storage facility site; 
\item[-] Tier 2 is external to the LOFAR LTA and consists of users who would like to retrieve data from 
the system and/or process the data outside the system. Additionally, all external astronomers and communities such as the Virtual Observatory are considered  
as Tier 2 users. 
\end{itemize}

Unlike the ``classic'' tier architecture of the LHC Computing Grid, the LOFAR Long-Term archive does not include data stored on 
Tier 0 and Tier 2 levels.

Two new systems were developed for the LTA: a data ingest system which allow to transfer data from Tier 0 to Tier 1 and an integrated Authentication and Authorization system which couples the A\&A systems used by the LOFAR observatory to the AstroWise A\&A.  

The data transfer from CEP to the archive includes the transfer of files with the data and ingestion of the metadata description of the data 
in the metadata database of the archive. The LOFAR data item consists of a number of files with observations, the data item is described 
in a single metadata file generated on Tier 0 by the MoM (Management of Measurements) system. 

In the case of the LOFAR the authorization of the user must be supervised by a system which is external to the archive, i.e., 
the LOFAR Authorization and Authentication system.

The LOFAR Observatory A\&A is the master which defines all users and their privileges  and  exports them to 
Astro-WISE A\&A. The use of a Novell Identity Manager (IdM) instance allows the synchronization of accounts bewteen the two systems.

\section{Current Status and Future Development}
\label{Future}

LTA is an example of the adaptation of the original Astro-WISE design to the additional requirements from the LOFAR community. The case of LOFAR shows the
high flexibility of Astro-WISE, which allows to include new elements keeping features of the original information system. 

LTA is a key project for Target\footnote{http://www.rug.nl/target} - the expertise center built
by the University of Groningen, Astron,
IBM and Oracle. Target develops information systems for a number of scientific communities including LOFAR. 

Target provides a storage facility of more than 10 PB capacity supporting the LOFAR LTA as well as a number of other applications. For processing purposes, the Groningen HPC cluster 
{\it milipede} is connected to the Target storage. The HPC cluster in Groningen has 250 nodes with at least 12 cores each and 24GB of memory on each node. The amount of local scratch 
space is 172GB per node. Next to the HPC cluster a new Grid cluster, financed by BiG Grid, will be installed and connected to the Target storage which provides about 60 compute nodes 
with 12-16 cores, 4GB of memory per core, and 3-4 TB of scratch space per system. Both clusters are part of Tier 1 and will be used for the LOFAR LTA data processing. 

\acknowledgements This work was performed as part of the Target project. Target project is supported by Samenwerkingsverband Noord Nederland. It operates under the 
auspices of Sensor Universe. It is also financially supported by the European fund for Regional Development and the Dutch Ministry of Economic Affairs, Pieken in de Delta, 
the Province of Groningen and the Province of Drenthe. LOFAR, the Low Frequency Array designed and constructed by ASTRON, has facilities in several countries, that are 
owned by various parties (each with their own funding sources), and that are collectively operated by the International LOFAR Telescope (ILT) foundation under a joint scientific policy.
This work is also supported by the programme of BiG Grid, the Dutch e-Science Grid, which is financially supported by the Netherlands Organisation for Scientific Research (NWO).

\bibliography{P009}

\end{document}